\documentclass[twocolumn]{revtex4}
\usepackage{graphicx}
\usepackage{dcolumn}
\usepackage{amsmath}

\begin{document}

\title{Crossover from weakly to strongly correlated regions in the
two-dimensional Hubbard model\\ 
-- Off-diagonal wave function Monte Carlo studies of Hubbard model II --
}

\author{Takashi \textsc{Yanagisawa}}

\affiliation{Electronics and Photonics Research Institute,
National Institute of Advanced Industrial Science and Technology (AIST),
Central 2, 1-1-1 Umezono, Tsukuba, Ibaraki 305-8568, Japan
}

\begin{abstract}
The ground state of the two-dimensional (2D) Hubbard model is investigated
by adopting improved wave functions that take into account intersite 
electron correlation beyond the Gutzwiller ansatz. 
The ground-state energy is lowered considerably, giving the best estimate
of the ground-state energy for the 2D Hubbard model.
There is a crossover from weakly to strongly correlated regions as
the on-site Coulomb interaction $U$ increases.
The antiferromagnetic correlation induced by $U$ is reduced for hole doping  
when $U$ is large, being greater than the bandwidth, thus increasing the kinetic
energy gain.
The spin and charge fluctuations are induced in the strongly correlated region.
These antiferromagnetic and kinetic charge fluctuations induce
electron pairings, which result in 
high-temperature superconductivity. 
\end{abstract}


\maketitle

\section{Introduction}
 
The mechanism of high-temperature superconductivity has been studied 
intensively since the 
discovery of cuprate high-temperature superconductors\cite{bed86,ben03}.
The correlation between electrons plays an important role 
because parent 
compounds without carriers are insulators.
It is primarily important to clarify the phase diagram of electronic states 
in the CuO$_2$ plane 
contained commonly in cuprate high-temperature superconductors.
It is obvious that an interaction with a large energy scale is 
necessary and responsible for the realization of high-temperature 
superconductivity.  The Coulomb interaction obviously has a large 
characteristic energy scale and can possibly be a candidate of the interaction that 
induces high-temperature superconductivity.

The CuO$_2$ plane consists of oxygen atoms and copper atoms.  
The electronic model for this plane is the d-p model (or three-band Hubbard 
model)\cite{eme87,hir89,sca91,web09,lau11,koi00,yan01,yan03,yan09,web14,koi03,koi06,koi01}.  
The single-band Hubbard model\cite{hub63,hub64,gut63} is obtained by neglecting 
oxygen atoms
in the CuO$_2$ plane.  
It is an open question whether the on-site Coulomb repulsion indeed 
induces superconductivity in correlated electron systems.  
This issue remains controversial for the two-dimensional (2D) Hubbard 
model\cite{aim07,zha97,zha97b,bul02,yan08,yan13}.
The studies on the ladder Hubbard model have indicated positive results for
superconductivity\cite{noa95,noa97,nak07,koi99,yam94,yan95}.  
These results suggest the existence of a pairing interaction affected
by the on-site Coulomb repulsive interaction.

A system described by the Hubbard model is a typical strongly correlated
system.  The 2D Hubbard model has been investigated
intensively for several decades\cite{hub63,hir85,yok88,whi89,loh90,gia91,mor92,yan96}.
The Hubbard model was first introduced to describe a metal--insulator
transition\cite{hub63}.
Since the discovery of cuprate high-temperature superconductors, many
researchers have tried to explain the occurrence of superconductivity in
cuprates in terms of the 2D Hubbard model.
The results of quantum Monte Carlo methods, which are considered to be
exact unbiased methods, do not support the existence of high-temperature 
superconductivity
in this model\cite{aim07,zha97,zha97b}.
In our opinion, this is because the Coulomb interaction $U$ is not large
in quantum Monte Carlo calculations, where the range of accessible $U$ is 
very restricted
because of the nature of the method.
On the basis of the variational Monte Carlo method, a finite superconducting
gap with $d$-wave symmetry is hardly obtained for $U/t < 6$ in the 2D Hubbard 
model\cite{nak97,yam98,yam00,yam11,yan13b,har09}.

It is necessary to improve the wave function in the variational Monte
Carlo method since the Gutzwiller function only accounts for the on-site
correlation.  The Gutzwiller function is a starting function that
should be improved to take account of correlation effects.
We discuss several methods of improving the wave function, and show that
an $\exp(-\lambda K)$-$P_G$-type function\cite{oht92,yan98,eic07} can be 
the best function with the lowest variational energy.
This wave function and its generalization were examined thoroughly on 
the basis of the variational Monte Carlo method in Ref. 45,
which is referred to as paper $I$ in this paper.
The variational energy is lowered considerably and
can be close to the exact value.

We discuss the properties of antiferromagnetism and superconductivity
in a strongly correlated
region by using improved wave functions.
It may be believed that the antiferromagnetic (AF) correlation is enhanced
as $U$ increases and that this will hold for large $U$ where $U$ is far greater
than the bandwidth.
This is, however, not true when holes are doped, as will be shown on the
basis of improved wave functions.  The AF correlation is
suppressed when $U$ is extremely large and larger than the bandwidth.
A superconducting correlation is developed in this region as the
AF correlation is suppressed with the increase in $U$.
The development of a superconducting correlation is understood to be induced by
spin fluctuation, which is induced by the kinetic charge fluctuation that
conquers antiferromagnetism.
This spin fluctuation in a strongly correlated region must be distinguished
from that in a weakly correlated region.  The latter is the conventional
spin fluctuation that has been discussed extensively in the
literature\cite{mor85,bic89}.

Thus, there is a crossover between weakly correlated and strongly
correlated regions as the Coulomb repulsion $U$ is varied.
A high-temperature superconductivity will be realized in the
strongly correlated region.

In the next section we discuss improved wave functions as below in correlated
electron systems.  In Sect. 3, we examine the stability of the 
AF and pairing states, on the basis of our wave functions.
We give a summary in the last section.
 
\section{Hamiltonian and Wave Functions}

\subsection{Two-dimensional Hubbard model}

The single-band Hubbard model is given by
\begin{equation}
H= \sum_{ij\sigma}t_{ij}c_{i\sigma}^{\dag}c_{j\sigma}
+U\sum_{i}n_{i\uparrow}n_{i\downarrow},
\end{equation}
where $t_{ij}$ are transfer integrals and $U$ is the on-site Coulomb
energy.  $n_{i\sigma}$ is the number operator given by
$n_{i\sigma}=c^{\dag}_{i\sigma}c_{i\sigma}$.  
The transfer integral $t_{ij}$ is nonzero,  $t_{ij}=-t$ for the
nearest-neighbor pair $\langle ij\rangle$ and $t_{ij}=-t'$ for
the next-nearest-neighbor pair $\langle\langle ij\rangle\rangle$.
Otherwise, $t_{ij}$ vanishes.  We denote the number of sites as $N$ and
the number of electrons as $N_e$.  The energy unit is given by $t$.
The second term represents the on-site Coulomb repulsive interaction
between electrons with opposite spins.
This simple term will illustrate profound phenomena in strongly correlated 
electron systems.

\subsection{Improved wave functions}

The wave function should include correlation between electrons.
A first step to include the electron correlation is 
the well-known Gutzwiller wave function, which reads as $\psi_G=P_G\psi_0$,
where $P_G$ is the Gutzwiller operator defined by
\begin{equation}
P_G= \prod_j\left( 1-(1-g)n_{j\uparrow}n_{j\downarrow}\right)
\end{equation}
with the variational parameter $g$ in the range of $0\le g\le 1$.
$P_G$ controls the double occupancy to take account of electron correlation. 
$\psi_0$ is a trial one-particle state that is taken to be the Fermi 
sea $\psi_{FS}$, the AF state (spin-density wave state)
$\psi_{AF}$, or the BCS state $\psi_{BCS}$.

The AF one-particle state $\psi_{AF}$ is given by the
eigenstate of the AF trial Hamiltonian:
\begin{equation}
H_{trial}= \sum_{ij\sigma}t_{ij}c^{\dag}_{i\sigma}c_{j\sigma}
-\Delta_{AF}\sum_{i\sigma}\sigma(-1)^{x_i+y_i}n_{i\sigma},
\end{equation}
where ${\bf r}_i\equiv (x_i,y_i)$ are the coordinates of site $i$.
$\Delta_{AF}$ indicates the AF order parameter.
In general, the transfer integrals $t_{ij}$ in the trial Hamiltonian
$H_{trial}$ can be treated as variational parameters so that the
ground-state energy is lowered.
In this paper, however, we do not use this procedure for simplicity,
and thus $t_{ij}$ in $H_{trial}$ are the same as those in the original
Hamiltonian.

One must improve the Gutzwiller wave function to lower the ground-state
energy because
only the on-site correlation is considered in the Gutzwiller ansatz.
It has been proposed that the wave function can be improved by taking account of 
the nearest-neighbor doublon-holon correlation\cite{kap82,yok06,miy12,yok13,sat16}.
The wave function with the doublon-holon correlation is given by
$\psi_{d-h}=P_{d-h}P_G\psi_0$ with
\begin{equation}
P_{d-h}=\prod_j\Big[ 1-(1-\eta)\prod_{\tau}[d_j(1-e_{j+\tau})
+e_j(1-d_{j+\tau})]\Big].
\nonumber\\
\end{equation}
Here, $d_j$ is the operator for the doubly occupied site given by
$d_j=n_{j\uparrow}n_{j\downarrow}$, and $e_j$ is the empty site operator
given as $e_j=(1-n_{j\uparrow})(1-n_{j\downarrow})$.
$\eta$ is a variational parameter in the range of $0\le\eta\le 1$.
We put $\eta=1$ in the non-interacting case.
A Jastrow factor is defined as
\begin{equation}
P_J = \exp\left( -\frac{1}{2}\sum_{i\neq j}h_{ij}n_in_j \right),
\end{equation}
where $n_i=n_{i\uparrow}+n_{i\downarrow}$ and $\{h_{ij}\}$ are 
variational parameters.
We can take into account intersite correlations by multiplying by $P_J$
to obtain wave functions such as $P_JP_{d-h}P_G\psi_0$.

A more efficient way to improve the wave function is to take account of 
the intersite correlation by
multiplying the Gutzwiller function by the kinetic operator.
A wave function of this type is written as\cite{oht92,yan98,yan99}
\begin{equation}
\psi_{\lambda}\equiv \psi^{(2)}=e^{-\lambda K}P_G\psi_0,
\end{equation}
where $K$ is the kinetic term in the Hamiltonian:
$K=\sum_{ij\sigma}t_{ij}c^{\dag}_{i\sigma}c_{j\sigma}$ and $\lambda$
is a variational parameter to be optimized to lower the energy.
The transfer integrals $t_{ij}$ in $K$ are not regarded as variational
parameters in this paper.
This wave function has been investigated by the variational Monte
Carlo method\cite{oht92,yan98,yan99,yan14} and a perturbative 
method\cite{eic07,eic09,bae09,bae11}.

This wave function can be further improved by multiplying by the
Gutzwiller operator and $e^{-\lambda K}$ again:
\begin{eqnarray}
\psi^{(3)}&\equiv& P_G\psi_{\lambda}= P_Ge^{-\lambda K}P_G\psi_0,\\
\psi^{(4)}&\equiv& e^{-\lambda' K}\psi^{(3)}= e^{-\lambda' K}P_G
e^{-\lambda K}P_G\psi_0.
\end{eqnarray}
The wave function $\psi^{(m)}$ is called the level-m wave function
in this paper.
This type of wave functions was first examined by the variational Monte
Carlo method in paper $I$\cite{yan98}.
The wave functions $\psi^{(2)}$, $\psi^{(3)}$, $\psi^{(4)}$,$\cdots$
contain intersite correlations of the site-off-diagonal type such as
$c^{\dag}_{i\sigma}c_{j\sigma}$, whereas the Gutzwiller and Jastrow
operators control only site-diagonal correlations such as 
$n_{i\sigma}n_{j\sigma'}$.

For the Gutzwiller-projected BCS-type wave function, we take $\psi_{BCS}$ 
as a trial one-particle state $\psi_0$:
\begin{equation}
\psi_{BCS}= \prod_k(u_k+v_kc^{\dag}_{k\uparrow}c^{\dag}_{-k\downarrow})
|0\rangle,
\end{equation} 
with coefficients $u_k$ and $v_k$ appearing only in the ratio
$v_k/u_k=\Delta_k/(\xi_k+\sqrt{\xi_k^2+\Delta_k^2})$, where $\Delta_k$ is the
${\bf k}$-dependent gap function and $\xi_k$ is the band dispersion
given as
\begin{equation}
\xi_{{\bf k}}= -2t(\cos k_x+\cos k_y)-2t'\cos k_x\cos k_y-\mu.
\end{equation}
We multiply by $P_{N_e}$ to extract only the state with the fixed total
electron number $N_e$. 
For the wave functions $\psi^{(2)}$, $\psi^{(3)}$,$\cdots$, it is
difficult to impose the site-diagonal-type constraint $P_{N_e}$ to
fix the total electron number.
To introduce the off-diagonal long-range order in the wave function with
off-diagonal site correlation, we perform the electron-hole
transformation for down-spin electrons\cite{yan99}:
\begin{equation}
d_k=c^{\dag}_{-k\downarrow},~~d^{\dag}_k=c_{-k\downarrow}.
\end{equation}
The up-spin electrons are unaltered, and we use the notation
$c_k=c_{k\uparrow}$.
If we denote the vacuum for $c$ and $d$ particles as $|\tilde{0}\rangle$,
for which $c_k|\tilde{0}\rangle=d_k|\tilde{0}\rangle=0$, the vacuum 
$|0\rangle$ reads as $|0\rangle=\prod_kd^{\dag}_k|\tilde{0}\rangle$.
Then, the BCS wave function is written as
\begin{equation}
\psi_{BCS} = \prod_k (u_k+v_kc^{\dag}_k d_k)|0\rangle 
= \prod_k (u_kd^{\dag}_k+v_kc^{\dag}_k)|\tilde{0}\rangle.
\end{equation}
In this wave function, $c$ and $d$ particles are mixed and the total
electron number is controlled by the chemical potential $\mu$.
We adjust the chemical potential so that we have the expectation value
$N_e$ for the total electron number.
 
In the above transformation, the Gutzwiller operator is mapped to
\begin{equation}
P_G= \prod_j (1-(1-g)c_j^{\dag}c_j(1-d_j^{\dag}d_j) ),
\end{equation}
and the Hamiltonian is transformed to
\begin{eqnarray}
H&=& -t\sum_{\langle ij\rangle}(c_i^{\dag}c_j-d_i^{\dag}d_j+{\rm h.c.})
\nonumber\\
&&-t'\sum_{\langle\langle j\ell\rangle\rangle}(c_j^{\dag}c_{\ell}
-d_j^{\dag}d_{\ell}+{\rm h.c.})+U\sum_ic_i^{\dag}c_i(1-d_i^{\dag}d_i),
\nonumber\\
\end{eqnarray}
where $c_j$ and $d_j$ are the Fourier transforms of $c_k$ and $d_k$,
respectively.
The averaged total electron number is written as
\begin{equation}
N_e = N+\sum_k\langle c_k^{\dag}c_k-d_k^{\dag}d_k\rangle.
\end{equation}

\subsection{Ground-state energy of improved wave functions}

We calculate the ground state energy to check the validity of our 
wave functions.
Recently, a variational wave function has been proposed by introducing 
a large number of variational parameters in the Gutzwiller--Jastrow
factor and the one-particle function\cite{tah08,mis14}.
The energy evaluated using a wave function with many variational parameters
is lowered considerably compared with that using the Gutzwiller function $\psi_G$.
Our wave function $\psi_{\lambda}$ also gives a good estimate of the
energy, and $P_G\psi_{\lambda}$ and $e^{-\gamma K}P_G\psi_{\lambda}$
give the best ground-state energy.
We show the variational energy calculated on a $4\times 4$ lattice in Table I
for a half-filled case and in Table II for hole doping.
The many-parameter wave function in Ref. 59 is a good wave function
with much lower variational energy.
The variational energy of our wave functions is also lowered considerably
compared with that of the Gutzwiller wave function and exhibits the best 
ground-state energy.
We show in Table III the expectation values of the spin correlation 
function $S({\bf q})$ for ${\bf q}=(\pi,\pi/2)$ on a $4\times 4$ lattice
with $N_e=12$, $t'/t=0$, and $U/t=10$.
The improved $\psi^{(2)}$ shows  good agreement with the exact value.

It is seen that the energy is not significantly improved only by multiplying the
Gutzwiller function by the doublon-holon correlation factor $P_{d-h}$ 
compared with that of the wave function with site-off-diagonal correlation.
The trial wave function $P_{d-h}P_G\psi_0$ was used to develop the 
physics of the Mott transition\cite{kap82} following the suggestion that
the Mott transition is due to doublon-holon binding\cite{mot49}.
We examined the Mott transition using the wave function
$e^{-\lambda K}P_G\psi_0$\cite{yan14} in a previous paper because the 
variational energy
of this wave function is 
much lower than that of the doublon-holon wave function.
The wave function $\psi_{\lambda}\equiv\psi^{(2)}$ clearly exhibited the 
Mott transition as $U$ increases.
 
The other way to improve wave functions is to regard the band parameters
in $\psi_0$ as variational parameters, that is, to take account of
the band renormalization.
There have been works in this direction\cite{miy12,sat16,him00,shi04}
We do not adopt this method in this paper to minimize the number of
variational parameters.
In Table IV, we show the ground-state energy to compare wave functions
on a $10\times 10$ lattice (which we mainly consider in this paper).
The level-2 wave function $\psi^{(2)}$ gives a good estimate of the
ground-state energy.

\begin{table}[htb]
\begin{center}
\caption{Variational energies for $4\times 4$ lattice, $N_e=16$,
$U/t=5$, and $t'/t=0$.  The boundary conditions are periodic in
one direction and antiperiodic in the other direction.
The nearest-neighbor (n.n.) Jastrow function in the third row means 
that we considered only
the nearest-neighbor Jastrow correlation factor
$\exp(-\sum_{\langle ij\rangle}h_{ij}n_in_j)$.
The one-particle state $\psi_{FS}$ indicates the Fermi sea.
The result in the fourth row is from Ref. 59.
The exact result was obtained by the exact diagonalization
method.
}
\begin{tabular}{lcl}\hline
Wave function  &  Energy  &  Comments \\ \hline
$P_G\psi_{FS}$  &  -11.654             &     \\
$P_{d-h}P_G\psi_{FS}$   &  -11.856      &  $g=0.46$, $\eta=0.89$ \\
$P_JP_{d-h}P_G\psi_{FS}$ & -11.863     & nearest neighbor Jastrow  \\
$P_JP_{d-h}P_G{\cal L}^{S=0}P_{pair}$ &  -12.459      &  Ref. 59 \\ 
$e^{-\lambda K}P_G\psi_{FS}$ & -12.366 & $g=0.15$, $\lambda=0.115$ \\
$P_G(g')e^{-\lambda K}P_G(g)\psi_{FS}$ & -12.479 & $g=0.035$, $\lambda=0.32$, $g'=0.66$ \\
$e^{-\lambda' K}P_Ge^{-\lambda K}P_G\psi_{FS}$ & -12.487 & $g=0.035$, $\lambda=0.32$, $g'=0.57$ \\
 & &  $\lambda'= 0.025$ \\
Exact  &  -12.530  &   \\ \hline
\end{tabular}
\end{center}
\end{table}

\begin{table}[htb]
\begin{center}
\caption{Variational energies for $4\times 4$ lattice, $N_e=12$,
and $t'/t=0$.  We chose $U/t=4$ and $U/t=10$.
The one-particle state $\psi_0$ is given by the Fermi sea, $\psi_0=\psi_{FS}$.
The boundary conditions are the same as in Table I.
}
\begin{tabular}{lccl}\hline
Wave function  &  $U/t=4$  &  $U/t=10$ & \\ \hline
$P_G\psi_{FS}$        & -18.239  & -13.940  &  $\psi_G=\psi^{(1)}$   \\
$P_{d-h}P_G\psi_{FS}$ & -18.266  & -14.024  &  \\
$P_JP_{d-h}P_G\psi_{FS}$ &  -18.269    & -14.031  & \\
$P_JP_{d-h}P_G{\cal L}^{S=0}P_{pair}$  & -18.406  & -14.435  &  Ref. 59 \\
$e^{-\lambda K}P_G\psi_{FS}$           & -18.481  & -14.544  & $\psi^{(2)}$ \\
$P_G(g')e^{-\lambda K}P_G(g)\psi_{FS}$ & -18.528  & -14.637  & $\psi^{(3)}$ \\
$e^{-\lambda' K}P_Ge^{-\lambda K}P_G\psi_{FS}$ & -18.536 & -14.685  & $\psi^{(4)}$ \\
Exact  &  -18.571  & -14.808 &  \\ \hline
\end{tabular}
\end{center}
\end{table}

\begin{table}[htb]
\begin{center}
\caption{Spin correlation function $S(\pi,\pi/2)$ for $4\times 4$ 
lattice with  $N_e=12$ and $t'/t=0$.
The parameters are $U/t=4$ and $U/t=10$.
We take the one-particle state $\psi_0$ as the Fermi sea, $\psi_0=\psi_{FS}$.
The last column shows $S(\pi,\pi/2)$ for $U/t=10$.
The boundary conditions are the same as in Table I.
}
\begin{tabular}{lccl}\hline
Wave function  &  $U/t=4$  &  $U/t=10$ & $(\pi,\pi/2)$  \\ \hline
$P_G\psi_{FS}$        & 0.0164  & 0.0206  &  0.0206    \\
$e^{-\lambda K}P_G\psi_{FS}$           & 0.0171  & 0.0220 & 0.0206   \\
$P_G(g')e^{-\lambda K}P_G(g)\psi_{FS}$ & 0.0177  & 0.0228 & 0.0209   \\
$e^{-\lambda' K}P_Ge^{-\lambda K}P_G\psi_{FS}$ & 0.0178 & 0.0221 & 0.0209  \\
$P_JP_{d-h}P_G{\cal L}^{S=0}P_{pair}$  & 0.0179  & 0.0261 &  \\
Exact  &  0.0179  & 0.0216 & 0.0216  \\ \hline
\end{tabular}
\end{center}
\end{table}

\begin{table}[htb]
\begin{center}
\caption{Variational energy per site on $10\times 10$ lattice
for comparison of wave functions.
The second column shows the ground energy for $U/t=12$,
$t'/t=-0.3$, and $N_e=92$.
The third column indicates the results for $U/t=18$,
$t'/t=0.0$, and $N_e=88$.
We take the one-particle state $\psi_0$ as the Fermi sea, $\psi_0=\psi_{FS}$,
or the AF state $\psi_0=\psi_{AF}$ in the second row.
The boundary conditions are periodic in one direction and
antiperiodic in the other direction.
The number in parentheses indicates the statistical error in last
digits.
}
\begin{tabular}{lccl}\hline
Wave function  &  $U=12$ $N_e=92$  &  $U=18$ $N_e=88$ & \\ \hline
$P_G\psi_{FS}$        & -0.3650(2)  & -0.4218(2)  & $\psi_G$    \\
$P_G\psi_{AF}$        & -0.3771(2)  & -0.4259(2)  &     \\
$P_{d-h}P_G\psi_{FS}$ & -0.4259(2)  & -0.4634(2)  &    \\
$P_JP_{d-h}P_G\psi_{FS}$ &  -0.4265(2)   & -0.4642(2)  & \\
$P_{d-h}P_G\psi_{AF-d}(BR)$  & -0.4915(2)  &  & Ref 53 \\
$e^{-\lambda K}P_G\psi_{FS}$   & -0.4956(2) & -0.5115(3)  & $\psi^{(2)}$ \\
$P_G(g')e^{-\lambda K}P_G(g)\psi_{FS}$ & -0.5095(3)  & -0.5175(3)  & $\psi^{(3)}$ \\
\hline
\end{tabular}
\end{center}
\end{table}

\section{Crossover from weakly to strongly correlated states}

\subsection{Ground-state energy}

We evaluate the ground-state energy of the 2D Hubbard
model with hole doping using $\psi_{\lambda}$, where
the one-particle state is taken as the Fermi sea, $\psi_0=\psi_{FS}$.
The details of the methods of Monte Carlo calculations with $\psi^{(m)}$ are 
shown in paper $I$.
We show the ground-state energy as a function of $U$ in Fig. 1,
where the calculations were performed on a $10\times 10$ lattice
with $N_e=88$ and $t'=0$. Solid circles indicate the energy obtained
by the Gutzwiller function and open circles denote that obtained using $\psi_{\lambda}$.
Squares represent the expectation values of the kinetic term 
(non-interacting part of the Hamiltonian) 
$E_K\equiv \langle\sum_{ij\sigma}t_{ij}c^{\dag}_{i\sigma}c_{j\sigma}\rangle$; 
solid squares
are for the Gutzwiller function and open squares are for $\psi_{\lambda}$.
The ground-state energy is lowered considerably by the optimization
with the $\lambda$ parameter.  The increase in kinetic energy gain is
appreciable, as shown in Fig. 1.
In this evaluation, we performed $5\times 10^4$ Monte Carlo steps with
about 200 parallel processors.

The double occupancy 
$\langle\sum_in_{i\uparrow}n_{i\downarrow}\rangle/N\equiv E_U/UN$ and
the expectation value of the Coulomb potential term 
$E_U/N\equiv \langle U\sum_in_{i\uparrow}n_{i\downarrow}\rangle/N$ are
shown in Fig. 2 and Fig. 3, respectively.
while the expectation value $E_U^G$ obtained by the Gutzwiller function is
larger than $E_U^{\lambda}$ obtained by using $\psi_{\lambda}$ in the weakly
correlated region, $E_U^{\lambda}$ becomes greater than $E_U^G$ for large $U$ 
in the strongly correlated region.
This suggests that the charge fluctuation is larger in $\psi_{\lambda}$
than in the Gutzwiller function.
To see the role of the parameter $\lambda$ in $\psi_{\lambda}$,
we show the energy as a function of $\lambda$ in Fig. 4.
The absolute value of $E_K$ increases as $\lambda$ increases, which 
lowers the total ground-state energy.
Thus, $\lambda$ induces charge fluctuation to increase the kinetic
energy gain.
This will bring about a crossover from weakly to strongly correlated
regions as $U$ increases.

\subsection{Crossover and antiferromagnetic correlation}

In this subsection, we investigate the stability of the antiferromagnetic state
as a function of $U$ in the case of hole doping.
The one-particle state is the AF state $\psi_0=\psi_{AF}$
with the order parameter $\Delta_{AF}$.
The AF correlation
is induced by the Coulomb repulsion $U$ and increases as $U$ increases.
In contrast, it is  suppressed if $U$ becomes larger than the bandwidth in 
the strongly correlated region.

Let us consider the spin correlation function $S({\bf q})$ given as
\begin{equation}
S({\bf q})= \frac{1}{4N}\sum_{ij}e^{i{\bf q}\cdot({\bf r}_i-{\bf r}_j)}
(n_{i\uparrow}-n_{i\downarrow})(n_{j\uparrow}-n_{j\downarrow}).
\end{equation}
$S({\bf q})$ has a maximum at ${\bf q}=(\pi,\pi)$.
The ${\bf q}=(\pi,\pi)$ component of $S({\bf q})$ is shown
in Fig. 5 as a function of $U$ for $N_e=88$ and $t'=0$ on a $10\times 10$
lattice, where the calculations are based on $\psi_{\lambda}$.
$S(\pi,\pi)$ has a peak near $U\simeq 10t$, showing the 
reduction in spin correlation for large $U$.
We show the antiferromagnetic (AF) order parameter $\Delta_{AF}$
as a function of $U$ in Fig. 6.
The AF order parameter is included in a trial wave function $\psi_0$
as a variational parameter.
The optimized $\Delta_{AF}$ shows a peak structure around $U\simeq 10t$
and vanishes in the strongly correlated region.
This is a typical behavior of $\Delta_{AF}$, and the AF energy gain
$\Delta E_{AF}$ also exhibits a similar behavior.
The calculation was carried out by employing the wave function $\psi_{\lambda}$
on a $10\times 10$ lattice.
When $U$ is small, $\Delta_{AF}$ increases with the increase in $U$ and
has a maximum at $U_{co}\simeq 10t$, which is on the order of the bandwidth.
When $U$ is larger than $U_{co}$, $\Delta_{AF}$ is decreased as $U$ is 
increased.  This indicates that AF correlation is suppressed for
extremely large $U$ and diminishes.  In the region $U>U_{co}$, there is competition
between the AF correlation and the charge fluctuation; this means that we must have
an AF energy gain or kinetic energy gain to lower the ground-state energy.
$\Delta_{AF}$ is reduced gradually as $U$ increases ($U>U_{co}$) since
the energy gain
$\Delta E_{AF}$ is presumably proportional to the AF exchange coupling,
$J\propto t^2/U$.
The AF correlation should be suppressed to obtain a kinetic energy gain
for large $U$.  Thus, we have a weak AF correlation in the strongly
correlated region with $U\ge U_{co}$.
This indicates that there is a large AF fluctuation in this region,
brought about by charge fluctuation,
where the charge fluctuation is driven
by the kinetic operator $K$ in the exponential factor $\exp(-\lambda K)$.
The correct term for the charge fluctuation is the kinetic charge 
fluctuation.

\subsection{Superconductivity in strongly correlated region}

It has been found that there is a large spin fluctuation being driven by 
the kinetic charge fluctuation.  We expect that a pairing interaction is 
induced by this kind of large spin and charge fluctuation.  

We perform calculations with the superconducting order parameter
in the wave function $\psi_{\lambda}$.
To do this, we use the electron-hole transformation for electrons
with the down spin as shown in the previous section.
The chemical potential is used to adjust the total electron number to
be equal to $N_e$.
We assume the $d$-wave symmetry for electron pairing and introduce
the order parameter $\Delta$.
A trial state is represented by a $2N\times N$ matrix $\phi$, and
the correlated pairing state can be formulated following the method
in Ref. 54.  We use a real--space formulation by using the
solution of the Bogoliubov--de Gennes equation\cite{yan99,miy04}: 
\begin{eqnarray}
\sum_j(H_{ij\uparrow}u_j^{\alpha}+F_{ij}v_j^{\alpha}v_j^{\alpha})
&=& E^{\alpha}u_i^{\alpha},\\
\sum_j(F_{ji}^*u_j^{\alpha}-H_{ji\downarrow}v_j^{\alpha})
&=& E^{\alpha}v_i^{\alpha},
\end{eqnarray}
for a trial Hamiltonian $H_{ij\sigma}$ and $F_{ij}$, where $(H_{ij\sigma})$
and $(F_{ij})$ are $N\times N$ matrices in the real space representation.
The matrix $(F_{ij})$ includes the gap function $\Delta$ as matrix
elements between $c$ and $d$ particles.
The initial matrix $\phi$ is given by $\phi_{i\alpha}=u_i^{\alpha}$
and $\phi_{i+N,\alpha}=v_i^{\alpha}$ for $i=1,\cdots,N$ and
$\alpha=1,\cdots,N$ with $E^{\alpha}>0$.

The energy plotted against $N_e/2$ is shown in Fig. 7 for $U/t=18$.
The figure indicates that the minimum of the
energy is at $\Delta/t\sim 0.06$, giving $E/N\simeq -0.5172$
 for $N_e=88$. 
In this case, the condensation energy per site is $\Delta E/N\simeq 0.0057t$.
The pairing state $e^{-\lambda K}P_G\psi_{BCS}$ with finite $\Delta$ is 
never stabilized 
for small $U$ such as $U<6t$.
The optimized superconducting order parameter increases as 
$U$ increases and has a maximum at some $U$ and greater than $U_{co}$.  
This is shown in Fig. 8, where the superconducting order parameter
$\Delta$ and the AF order parameter are shown as
functions of $U$.

It is appropriate to call the region $U>U_{co}$ the strongly correlated 
region because the ground state at half-filling is insulating in this
region\cite{yok06,yan14}.  
It is difficult to find a clear sign of superconductivity in weakly correlated 
regions, where $U\le 6t$, using numerical methods such as a quantum
Monte Carlo method\cite{aim07,zha97,yan13}.
This is also the case for the variational Monte Carlo method since the
superconducting condensation energy vanishes or is very small for small $U$.
Instead, in a strongly correlated region, we obtain a clear indication
of superconductivity.

\begin{figure}
\begin{center}
\includegraphics[width=8.0cm]{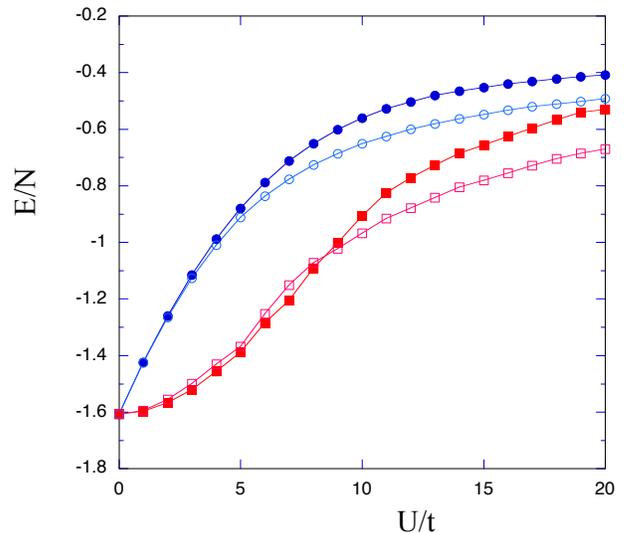}
\caption{
Ground-state energy as a function of
$U$ in units of $t$ on $10\times 10$ lattice.
The number of electrons is $N_e = 88$ and we set $t'= 0.0$.
We used the periodic boundary condition in one direction and the antiperiodic
one in the other direction.
Solid symbols are obtained using the Gutzwiller function and open
symbols are obtained using the improved function $\psi_{\lambda}$ with the optimized
parameter $\lambda$.
The circles and squares indicate the ground-state energy and kinetic
energy, respectively. 
}
\end{center}
\label{E-88}
\end{figure}

\begin{figure}
\begin{center}
\includegraphics[width=8.0cm]{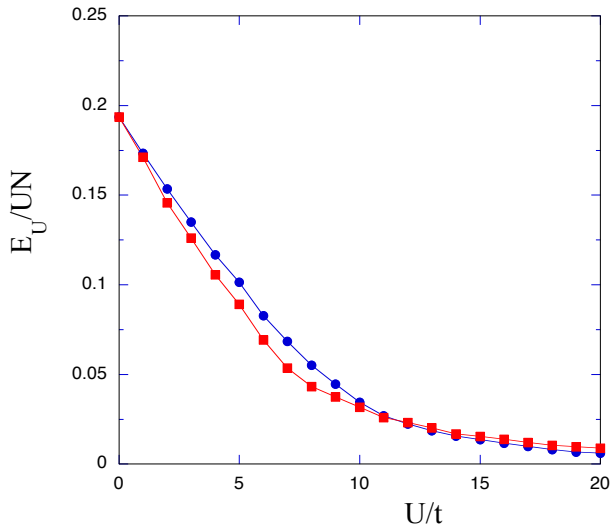}
\caption{
Double occupancy $\langle\sum_in_{i\uparrow}n_{i\downarrow}\rangle/N$ as 
a function of $U$
on $10\times 10$ lattice.
The circles are for the Gutzwiller function and the squares are for 
$\psi_{\lambda}$.
The number of electrons is $N_e = 88$ and we set $t'= 0.0$.
The boundary conditions are the same as in Fig. 1
In the small-$U$ region, the double occupancy of the Gutzwiller function
is greater than that of $\psi_{\lambda}$, and this is reversed
in the large-$U$ region.
}
\end{center}
\label{dbl-88}
\end{figure}

\begin{figure}
\begin{center}
\includegraphics[width=8.0cm]{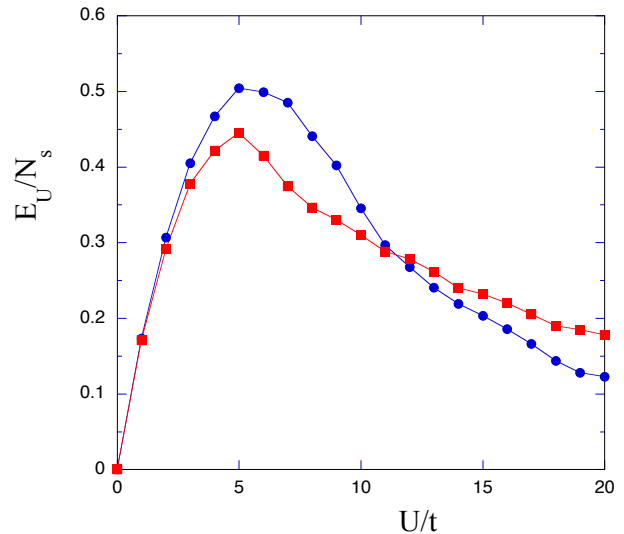}
\caption{
Expectation value of the Coulomb term 
$E_U\equiv \langle U\sum_in_{i\uparrow}n_{i\downarrow}\rangle$ as a 
function of $U$
on $10\times 10$ lattice.
The number of electrons is $N_e = 88$ and we set $t'= 0.0$.
The boundary conditions are the same as in Fig. 1
The expectation value of the Coulomb interaction obtained by the Gutzwiller
function $E_U^G$ (indicated by circles) is larger than the expectation value 
$E_U^{\lambda}$ (squares) obtained using
$\psi_{\lambda}$ for small $U$.  For large $U$, we have 
$E_U^{\lambda}>E_U^G$.
}
\end{center}
\label{UE-88}
\end{figure}

\begin{figure}
\begin{center}
\includegraphics[width=8.0cm]{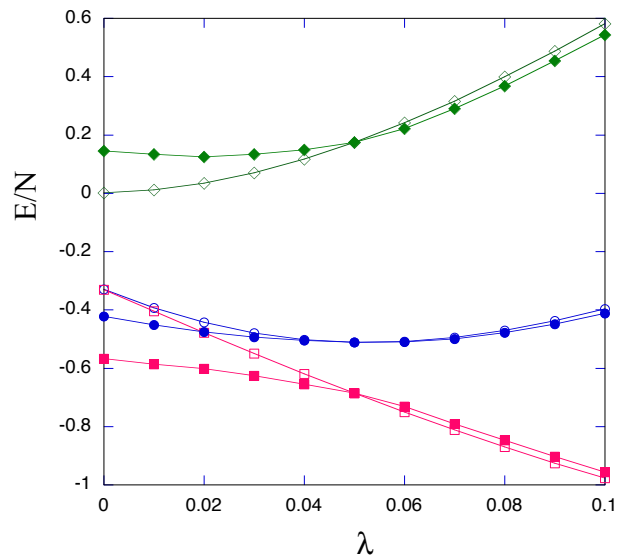}
\caption{
Ground-state energy (circles), kinetic energy (squares), and
Coulomb energy $E_U/N$ (diamonds) as functions of $\lambda$
on $10\times 10$ lattice for $U/t=18$.
The number of electrons is $N_e = 88$ and we set $t'= 0.0$.
The boundary conditions are the same as in Fig. 1
Open symbols indicate the values obtained with the variational
parameter $g$ fixed to the optimized value at which the total
energy is minimized.  For solid symbols, $g$ is also optimized. 
As is clear in the figure, the kinetic energy gain increases as
$\lambda$ increases.  This lowers the ground-state energy.
}
\end{center}
\label{lambda-88}
\end{figure}

\begin{figure}
\begin{center}
\includegraphics[width=8.5cm]{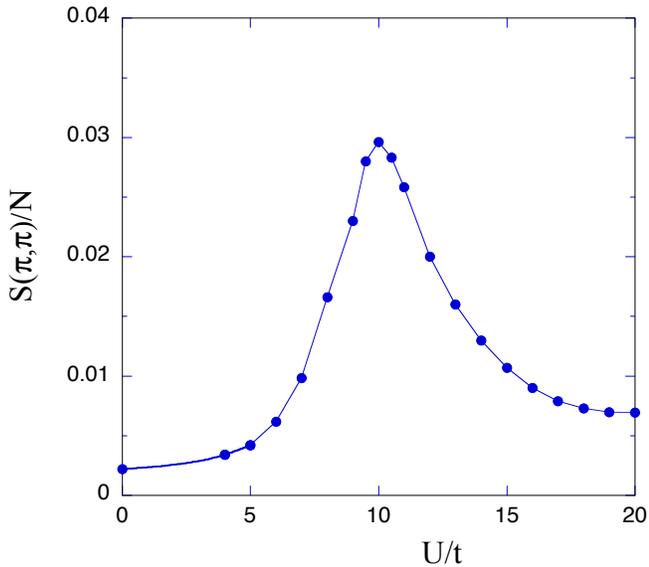}
\caption{
Spin correlation function $S(\pi,\pi)$ as a function of $U$
on $10\times 10$ lattice.
The number of electrons is $N_e = 88$ and we set $t'= 0.0$.
The boundary conditions are the same as in Fig. 1.
}
\end{center}
\label{sqm-88}
\end{figure}

\begin{figure}
\begin{center}
\includegraphics[width=8.5cm]{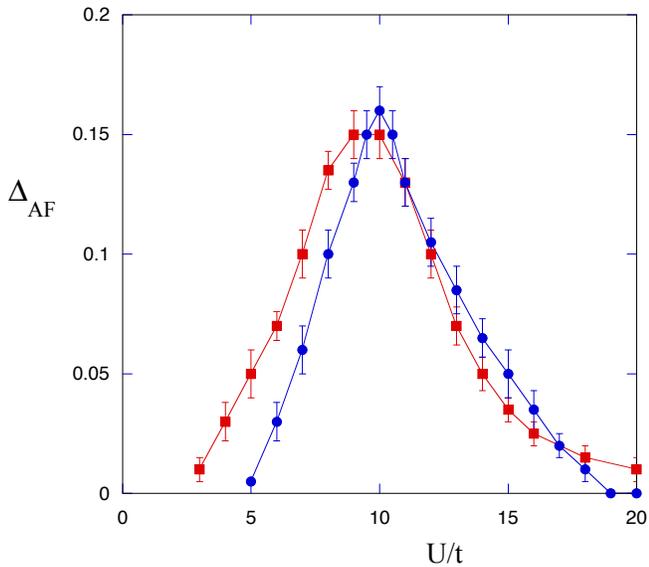}
\caption{
Antiferromagnetic order parameter $\Delta_{AF}$ as a function of
$U$ on $10\times 10$ lattice.
The number of electrons is $N_e=88$ (squares) for $t'=0$ with the same
boundary conditions as in Fig. 1. 
We also show the results for $N_e = 84$ with $t'=-0.2t$ (circles), where
we used the periodic boundary condition in both directions.
}
\end{center}
\label{100-afu}
\end{figure}

\begin{figure}
\begin{center}
\includegraphics[width=8.5cm]{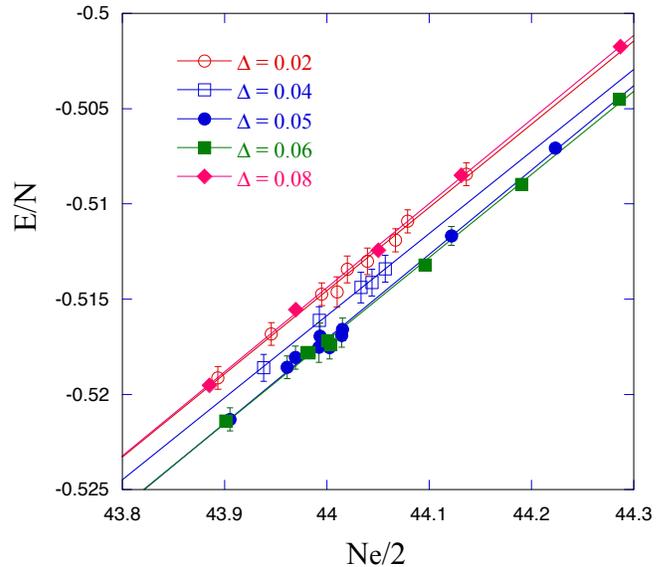}
\caption{
Ground-state energy as a function of the electron number $N_e/2$
for fixed superconducting gap $\Delta$
on $10\times 10$ lattice.
The number of electrons is $N_e = 88$, and we set $U=18t$ and $t'= 0.0$.
The energy unit is given by $t$.
The boundary conditions are the same as in Fig. 1.
}
\end{center}
\label{100-88e-ne}
\end{figure}

\begin{figure}
\begin{center}
\includegraphics[width=8.5cm]{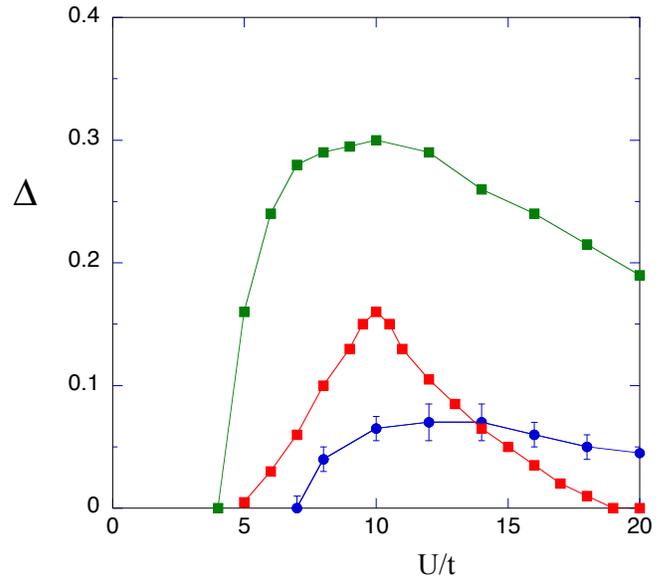}
\caption{
Superconducting and antiferromagnetic order parameters as functions of
of $U$ in units of $t$ on $10\times 10$ lattice.
The number of electrons is $N_e = 88$ and $t'=0.0$.
The solid circles show the SC gap for the improved function.
The squares represent the antiferromagnetic order parameter, where
the upper curve is for the Gutzwiller function and the lower curve
is for the improved function.
The boundary condition is periodic in one direction and antiperiodic
in the other direction.
}
\end{center}
\label{100-88sc}
\end{figure}

\section{Summary}

We have investigated the 2D Hubbard model by adopting 
improved wave functions 
that take into account intersite correlations beyond the Gutzwiller ansatz.  
Our wave function is an $\exp(-\lambda K)$-$P_G$-type wave function, which is 
inspired
by the wave function used in quantum Monte Carlo methods and
can be improved systematically by multiplying by $P_G$ and
$e^{-\lambda K}$.
We can improve the variational energy considerably with this wave function.
Our wave function can give the best estimate of the ground-state energy of
the 2D Hubbard model.

The improved wave function gives results that are qualitatively
different from those obtained by the simple Gutzwiller function.
In particular, the picture of the stability of the antiferromagnetic state 
is crucially changed when we employ wave functions with intersite
correlations.
The antiferromagnetic state is stable when $U$ is larger than a critical
value at half-filling.  Although one may think that this also holds for the
hole-doped case, this is not true away from half-filling.  
To obtain the kinetic energy gain, we must eliminate the antiferromagnetic order
when $U$ is extremely large because the energy lowering by the antiferromagnetic
order is small, being proportional to $t^2/U$.
Thus, the antiferromagnetic correlation will fade away as $U$ increases
to be greater than the critical $U_{co}$.

The reduction in the antiferromagnetic correlation in the strongly correlated 
region suggests the existence of a large antiferromagnetic spin fluctuation.  
The charge fluctuation induced by the kinetic operator is 
appreciable and helps electrons to form pairs.
We expect that these large spin and charge fluctuations induce an
effective pairing interaction 
between electrons, which may be able to bring about high-temperature 
superconductivity.  There is also spin 
fluctuation in the weakly correlated region, where the antiferromagnetic 
correlation fades away as $U$ is reduced.  

The next-nearest-neighbor transfer $t'/t$ is expected to be important in 
determining the stability of the antiferromagnetic state.  It has been argued
that the ground-state property, especially the stability of ordered states, 
crucially depends on $t'/t$\cite{sat16,tah08} and that the antiferromagnetic
state will be stabilized due to $t'/t$.
This will be an interesting future subject when studied on the basis of our improved
wave functions.  

A discussion on the relationship between the t-J model and the
Hubbard model is given here.
It is well known that the energy scale is given by $J=4t^2/U$ in the
large-$U$ region, as shown by mapping the Hubbard model to a
t-J-like model\cite{har67}.
The $U$-dependence of the antiferromagnetic correlation can be
understood from this mapping, namely, the AF correlation is suppressed in the
large-$U$ region such as $U> 8t$.
This suggests that the physics in the strongly correlated region is
similar to that of the t-J model.
In the t-J model, the exchange interaction $J$ induces superconductivity
with $d$-wave pairing as well as AF ordering\cite{yok96}.
At half-filling, the ground state has an AF long-range order for $J>0$.
This AF ordering shows instability due to hole doping; this occurs even for
small doping.  In the case of the Hubbard model for large $U$, say $U\sim 20t$,
the AF state becomes unstable upon hole doping.
This indicates a similarity between the t-J model and the Hubbard model in the
strongly correlated region.
There is, however, a difference between the two models.
For example, the superconducting gap $\Delta$ remains finite even for large $U$
with small $J$, where
the AF order is suppressed.
This means that we cannot well understand $\Delta$ as a function of $U$ only 
by means of $J$.
We consider that we must take account of the charge fluctuation to understand
the superconducting state for large $U$.
We must acknowledge that fluctuations are important, and in particular, the
charge fluctuation, as well as spin fluctuation, plays a significant role
in stabilizing the pairing state.
 
The crossover behavior from weakly to strongly correlated regions
or vice versa in the 2D Hubbard model may belong to a
universality class.
The well-known Kondo problem exhibits a crossover from the weak 
coupling region to the strong coupling region as the temperature $T$
decreases\cite{kon12}.
The s-d model is mapped to an electron-gas model interacting with
a logarithmic interaction, from which the scaling equation
for the coupling constant $J$ was derived by the renormalization
group procedure\cite{yuv70,yuv70b}.
The renormalization group equation for the s-d model agrees with that for the
2D sine-Gordon model\cite{kog79,ami80,yan16}.
The one-dimensional (1D) Hubbard model is mapped onto the 2D sine-Gordon
model by a bosonization procedure\cite{sol79,hal81}.
Thus, these models (2D sine-Gordon model, s-d model, and 1D Hubbard model) 
are in the same universality class from the viewpoint of scaling theory.
We expect that, concerning the crossover between weakly and strongly 
correlated regions, the 2D Hubbard model is also in this class
given by the 2D sine-Gordon model.

In conclusion, the crossover occurs from a weakly correlated to a
strongly correlated region, and high-temperature superconductivity
will appear in the strongly correlated region.

\section*{Acknowledgments}
The author expresses his sincere thanks to K. Yamaji, M. Miyazaki, I. Hase,
and S. Koikegami for valuable discussions.
The computations were supported by the Supercomputer
Center of the Institute for Solid State Physics, The University of Tokyo.
This work was supported by a Grant-in-Aid for Scientific Research from the
Ministry of Education, Culture, Sports, Science and Technology of Japan
(Grant No. 22540381).


\vspace{1cm}

\begin{thebibliography}{9}

\bibitem{bed86} J. G. Bednorz and K. A. M\"{u}ller, Z. Phys. B64, 189 (1986).
\bibitem{ben03} {\em The Physics of superconductors}, 
edi. K. H. Bennemann and J. B. Ketterson (Springer-Verlag, Berlin,
2003), Vols. I and II.
\bibitem{eme87} V. J. Emery, Phys. Rev. Lett. 58, 2794 (1987).
\bibitem{hir89}
J. E. Hirsch, D. Loh, D. J. Scalapino, and S. Tang, Phys. Rev. B39, 243 (1989).
\bibitem{sca91}
R. T. Scalettar, D. J. Scalapino, R. L. Sugar, and S. R. White, Phys. Rev. B44,
770 (1991).
\bibitem{web09}
C. Weber, A. Lauchi, F. Mila, and T. Giamarchi, Phys. Rev. Lett. 102, 017005 (2009).
\bibitem{lau11}
B. Lau, M. Berciu, and G. A. Sawatzky, Phys. Rev. Lett. 106, 036401 (2011).
\bibitem{koi00} S. Koikegami and K. Yamada, J. Phys. Soc. Jpn. 69, 768 (2000).
\bibitem{yan01}
T. Yanagisawa, S. Koike, and K. Yamaji, Phys. Rev. B64, 184509 (2001).
\bibitem{yan03}
T. Yanagisawa, S. Koike, and K. Yamaji, Phys. Rev. B67, 132408 (2003).
\bibitem{yan09}
T. Yanagisawa, M. Miyazaki, and K. Yamaji, J. Phys. Soc. Jpn. 78, 013706 (2009).
\bibitem{web14} C. Weber, T. Giamarchi, and C. M. Varma, Phys. Rev. Lett. 
112, 117001 (2014).
\bibitem{koi03}
S. Koikegami and T. Yanagisawa, Phys. Rev. B67, 134517 (2003).
\bibitem{koi06}
S. Koikegami and T. Yanagisawa, J. Phys. Soc. Jpn. 75, 034715 (2006).
\bibitem{koi01}
S. Koikegami and T. Yanagisawa, J. Phys. Soc. Jpn. 70, 3499 (2001).

\bibitem{hub63}J. Hubbard, Proc. Roy. Soc. London 276, 238 (1963).
\bibitem{hub64}J. Hubbard, Proc. Roy. Soc. London 281, 401 (1964).
\bibitem{gut63}M. C. Gutzwiller, Phys. Rev. Lett. 10, 159 (1963).

\bibitem{aim07}
T. Aimi and M. Imada, J. Phys. Soc. Jpn. 76, 113708 (2007).
\bibitem{zha97}
S. Zhang, J. Carlson, and J. E. Gubernatis, Phys. Rev. B55, 7464 (1997).
\bibitem{zha97b}
S. Zhang, J. Carlson, and J. E. Gubernatis, Phys. Rev. Lett. 78, 4486 (1997).
\bibitem{bul02}
N. Bulut, Advances in Phys. 51, 1587 (2002).
\bibitem{yan08}
T. Yanagisawa, New J. Phys. 10, 023014 (2008).
\bibitem{yan13}
T. Yanagisawa, New J. Phys. 15, 033012 (2013).
\bibitem{noa95} R. M. Noack, S. R. White, and D. J. Scalapino, EPL 30, 163 (1995).
\bibitem{noa97}
R. M. Noack, N. Bulut, D. J. Scalapino, and M. G. Zacher, Phys. Rev. B56, 7162 (1997).
\bibitem{nak07} T. Nakano, K. Kuroki, and S. Onari, Phys. Rev. B76, 014515 (2007).
\bibitem{koi99}
S. Koike, K. Yamaji, and T. Yanagisawa, J. Phys. Soc. Jpn. 68, 1657 (1999).
\bibitem{yam94}
K. Yamaji, Y. Shimoi, and T. Yanagisawa, Physica C235-240, 2221 (1994).
\bibitem{yan95}
T. Yanagisawa, Y. Shimoi, and K. Yamaji, Phys. Rev. B52, R3860 (1995).

\bibitem{hir85}
J. E. Hirsch, Phys. Rev. B31, 4403 (1985).
\bibitem{yok88}
H. Yokoyama and H. Shiba, J. Phys. Soc. Jpn. 57, 2482 (1988).
\bibitem{whi89}
S. R. White, D. J. Scalapino, R. L. Sugar, E. Y. Loh, J. E. Gubernatis, and
R. T. Scalettar, Phys. Rev. B40, 506 (1989).
\bibitem{loh90} E. Y. Loh, J. E. Gubernatis, R. T. Scalettar, S. R. White,
D. J. Scalapino, and R. L. Sugar, Phys. Rev. B41, 9301 (1990).
\bibitem{gia91} T. Giamarchi and C. Lhuillier, Phys. Rev. B43, 12943 (1991).
\bibitem{mor92}
A. Moreo, Phys. Rev. B45, 5059 (1992).
\bibitem{yan96}
T. Yanagisawa and Y. Shimoi, Int. J. Mod. Phys. B10, 3383 (1996).
\bibitem{nak97}
T. Nakanishi, K. Yamaji, and T. Yanagisawa, J. Phys. Soc. Jpn. 66, 294 (1997).
\bibitem{yam98}
K. Yamaji, T. Yanagisawa, T. Nakanishi, and S. Koike, Physica C304, 225 (1998).
\bibitem{yam00}
K. Yamaji, T. Yanagisawa, and S. Koike, Physica B284-288, 415 (2000).
\bibitem{yam11}
K. Yamaji, T. Yanagisawa, M. Miyazaki, and R. Kadono, J. Phys. Soc. Jpn. 
80, 083702 (2011).
\bibitem{yan13b}
T. Yanagisawa, M. Miyazaki, and K. Yamaji, J. Mod. Phys. 4, 33 (2013).
\bibitem{har09}
T. M. Hardy, J. P. Hague, J. H. Samson, and A. S. Alexandrov,
Phys. Rev. B79, 212501 (2009).
\bibitem{oht92} H. Otsuka, J. Phys. Soc. Jpn. 61, 1645 (1992).
\bibitem{yan98}
T. Yanagisawa, S. Koike, and K. Yamaji, J. Phys. Soc. Jpn. 67, 3867 (1998).
\bibitem{eic07} D. Eichenberger and D. Baeriswyl, Phys. Rev. B76, 180504 (2007).
\bibitem{mor85} T. Moriya, {\em Spin Fluctuations in Itinerant
Electron Magnetism} (Springer, Berlin, 1985).
\bibitem{bic89} N. E. Bickers, D. J. Scalapino, and S. R. White,
Phys. Rev. Lett. 62, 961 (1989).

\bibitem{kap82} T. A. Kaplan, P. Horsh, and P. Fulde, Phys. Rev. Lett.
49, 889 (1982).
\bibitem{yok06}
H. Yokoyama, M. Ogata and Y. Tanaka, J. Phys. Soc. Jpn. 75, 114706 (2006).
\bibitem{miy12} M. Miyazaki, T. Yanagisawa, and K. Yamaji, Phys. Procedia
27, 64 (2012).
\bibitem{yok13}
H. Yokoyama, M. Ogata, Y. Tanaka, K. Kobayashi, and H. Tsuchiura,
J. Phys. Soc. Jpn. 82, 014707 (2013).
\bibitem{sat16} R. Sato and H. Yokoyama, J. Phys. Soc. Jpn. 85, 074701
(2016).
\bibitem{yan99}
T. Yanagisawa, S. Koike, and K. Yamaji, J. Phys. Soc. Jpn. 68, 3608 (1999).
\bibitem{yan14}
T. Yanagisawa and M. Miyazaki, EPL 107, 27004 (2014).
\bibitem{eic09} E. Eichenberger and D. Baeriswyl, Phys. Rev. B79, 100510 (2009).
\bibitem{bae09} D. Baeriswyl, D. Eichenberger, and M. Menteshashvii, New J.
Phys. 11, 075010 (2009).
\bibitem{bae11} D. Baeriswyl, J. Supercond. Novel Magn. 24, 1157 (2011).

\bibitem{tah08} D. Tahara and M. Imada, J. Phys. Soc. Jpn. 77, 114701 (2008).
\bibitem{mis14} T. Misawa and M. Imada, Phys. Rev. B90, 115137 (2014).
\bibitem{mot49} N. F. Mott, Proc. Phys. Soc., Sect. A62, 416 (1949).
\bibitem{him00} A. Himeda and M. Ogata, Phys. Rev. Lett. 85, 4345 (2000).
\bibitem{shi04} C. T. Shih, T. K. Lee, R. Eder, C.-Y. Mou and
Y. C. Chen, Phys. Rev. Lett. 92, 227002 (2004).
\bibitem{miy04}
M. Miyazaki, T. Yanagisawa and K. Yamaji, J. Phys. Soc. Jpn. 73, 1643 (2004).

\bibitem{har67} A. B. Harris and R. V. Lange, Phys. Rev. B157, 295 (1967).
\bibitem{yok96} H. Yokoyama and M. Ogata, J. Phys. Soc. Jpn. 65, 3615 (1996).
\bibitem{kon12}J, Kondo, {\em The Physics of Dilute Magnetic Alloys}
(Cambridge University Press, Cambridge, 2012).
\bibitem{yuv70} G. Yuval and P. W. Anderson, Phys. Rev. B1, 1522 (1970).
\bibitem{yuv70b} P. W. Anderson, G. Yuval, and D. R. Hamann, Phys. Rev. 
B1, 4464 (1970).
\bibitem{kog79} J. B. Kogut, Rev. Mod. Phys. 51, 659 (1979).
\bibitem{ami80} D. J. Amit, Y. Y. Goldschmidt, and S. Grinstein,
J. Phys. A: Math. Gen. 13, 585 (1980).
\bibitem{yan16} T. Yanagisawa, EPL 113, 41001 (2016).
\bibitem{sol79} J. S\'{o}lyom, Adv. Phys. 28, 201 (1979).
\bibitem{hal81} F. D. N. Haldane, J. Phys. C14, 2585 (1981).




\end{thebibliography}
\end{document}